# A Bayesian Mixture Model for Changepoint Estimation Using Ordinal Predictors

Emily Roberts, MS (ORCID 0000-0002-5838-9691) and Lili Zhao, PhD
Department of Biostatistics, University of Michigan
August 1, 2020

## ABSTRACT

In regression models, predictor variables with inherent ordering, such as tumor staging ranging and ECOG performance status, are commonly seen in medical settings. Statistically, it may be difficult to determine the functional form of an ordinal predictor variable. Often, such a variable is dichotomized based on whether it is above or below a certain cutoff. Other methods conveniently treat the ordinal predictor as a continuous variable and assume a linear relationship with the outcome. However, arbitrarily choosing a method may lead to inaccurate inference and treatment. In this paper, we propose a Bayesian mixture model to simultaneously assess the appropriate form of the predictor in regression models by considering the presence of a changepoint through the lens of a threshold detection problem. By using a mixture model framework to consider both dichotomous and linear forms for the variable, the estimate is a weighted average of linear and binary parameterizations. This method is applicable to continuous, binary, and survival outcomes, and easily amenable to penalized regression. We evaluated the proposed method using simulation studies and apply it to two real datasets. We provide JAGS code for easy implementation.

## ACKNOWLEDGEMENTS/SOURCE OF FUNDING

*This work was supported by the* National Science Foundation (DGE 1256260) to E. Roberts; the NIH (P30 CA 046592-28) to L. Zhao.

Corresponding author information

Emily Roberts, Department of Biostatistics, University of Michigan, 1415 Washington Heights, Ann Arbor, MI 48109, email: ekrobe@umich.edu

## 1. INTRODUCTION

Qualitative variables with a logical ordering are frequently seen in medical research, such as tumor staging ranging from zero to four and ECOG performance status ranging from zero to five. This paper is motivated by two projects involving the study of effects of ordinal predictors on patient outcomes. The first is the preoperative diagnosis of thyroid malignancies. Fine-needle aspiration (FNA) of the thyroid is a common practice in the management of thyroid nodules.[1] Recognition of various architectural patterns and cytologic features in nodules is essential in distinguishing neoplastic nodules, which require surgical intervention, from non-neoplastic nodules, which may be conservatively managed by clinical and radiologic follow-up. The cytologic features are graded by a pathologist on the scale of 0-3 indicating absent, noticeable, easily seen, and prominent. We want to learn which feature(s) can be used to predict the probability of malignancy. The second project is determining potential relationships between tumor expression data from The Cancer Genome Atlas and overall survival time. We are interested in associating the ordinal protein expression data with patient survival outcomes.

In studying the association of these ordinal predictors with patient outcomes, we face two decisions: either treating the predictors as a continuous or a categorical variable. If we treat the ordinal predictor as a continuous variable, we can assume a linear relationship with the outcome with an appropriate link function among the ordered levels. The linear assumption is convenient and probably sufficient when the ordinal predictor only has a few (around four or five) ordered categories and there is insufficient information to construct a regression model with higher orders. However, this assumption requires that the effect on the outcome is the same between consecutive levels. If the researcher can justify this assumption, treating the variable as numeric will achieve the greatest power and make best use of the data information. However, if this linear assumption does not fit the true behavior of the variable, the regression coefficients under this assumption will provide an incorrect interpretation.[2]

To avoid the linearity assumption, a common strategy is to treat the ordinal predictor as a categorical variable with *k* levels. In this approach, each category is treated separately, allowing for different magnitudes of changes for consecutive categories. When the predictor variable only has a few categories as in our two projects, one strategy is to dichotomize the ordinal predictor into a binary variable using a cutoff (i.e., low vs high values). This approach is common in practice because 1) the treatment decision is generally based on whether the ordinal predictor is above or below the threshold, and 2) the small number of subjects in some categories do not allow us to estimate two or more cutoffs. Various statistical approaches have been developed to find the optimal cutoff for the dichotomization. Repeated or arbitrary methods that create many possible thresholds for a cutoff to test for predictive ability or the lowest p-value are discouraged and have been criticized for their ad-hoc nature, potential for overfitting, and lack of correction for multiple comparison.[3-5] However, converting the ordinal variable into a binary variable may not make use of all data information.

Often an increased value of a risk factor corresponds to a greater likelihood of disease with a monotonic relationship between risk and successive groups, while in other instances, there exists a changepoint where the effect of the predictor differs suddenly.[6] Examples of this phenomenon and corresponding models have been studied extensively for different scenarios.[7] A physician may rely on detecting this change to give the most effective treatment. However, it may be unclear how to determine the cutoff. The physician may use previous experience to determine a clinically significant threshold, but a more formal statistical model may be preferable as arbitrarily choosing a method may lead to inaccurate inference and treatment.

To compromise between assuming that a linear relationship exists among levels of an ordinal predictor and treating each level independently, advanced methods have been published. Latent variables provide a framework to model ordinal variables by assuming an underlying continuous process exists and follows some distributional assumptions.[8-9] If this continuous assumption does not hold, it is also possible to transform the predictor in some way to be linear, but appropriate transformations may be difficult to determine or interpret. Another approach is isotonic regression, which can be used to preserve ordering in variables by enforcing a regression line to be non-decreasing between categories.[10-13] However, the direction of the monotonic relationship (either non-decreasing or non-increasing between categories) needs to be pre-specified before applying isotonic regression. This is very challenging when the model has many ordinal predictors and the directions of association for the predictors are unknown.

In this paper, we develop a simple approach that assumes that the ordinal predictor either has a linear relationship with the outcome (via a link function), or the effect of the ordinal predictor can be expressed as high versus low. Our outcome of interest may take several forms such as continuous, binary (neoplasia vs non-neoplasia in the case of the FNA example), or survival (time-to-death or progression-free survival). Linear, logistic, and Cox proportional hazards models are common regression models for each of these respective data types with no true consensus on how to analyze ordinal predictors effectively. We will show that our approach is very simple to implement with only a few lines of code and has a straightforward interpretation. Furthermore, it is easily amenable to perform variable selection using penalized regression.

In section 2, we propose a statistical method to determine if a changepoint exists in the effect of an ordinal variable, or if, in the absence of such a threshold, it should be treated linearly. We use these two functions to form the basis of our model. If a threshold truly exists, the model can detect the optimal cutoff value for the ordinal variable and simultaneously estimate its significance. By allowing for the possibility that the ordinal input variable should be treated as continuous, the model avoids the risk of distorting estimation and inference through dichotomization. We demonstrate the performance of the method via simulation in section 3 and with two data examples in section 4. We conclude with final remarks in section 5. Code for implementation can be found in the appendix (available upon request).

## 2. METHODS

2.1 Regression models

In this section, we construct a general regression model allowing for a continuous, binary or survival outcome.

Let $y_i$ denote the outcome measured for subject $i$ ($i = 1, \ldots, n$) and $x_{ij}$ denote the $j$th ordinal predictor variable ($j = 1, \ldots, J$) taking a value 0, 1, …, or K. We model $y_i$ as a function of the $J$ predictors. If the outcome $y_i$ is continuous, we have

$$y_i \sim N(\mu_i, \sigma^2) \qquad (1)$$

where $\mu_i = \alpha + \beta_1 f(x_{i1}) + \cdots + \beta_j f(x_{ij}) + \cdots + \beta_J f(x_{iJ})$, and $f(x_{ij})$ is defined as

$$f(x_{ij}) = I(Z_j = 1) \cdot \frac{x_{ij}}{2 \cdot sd(x_j)} + I(Z_j = 0) \cdot I(x_{ij} < \tau_j)$$

The latent parameter $Z_j$ takes a value 0 or 1 representing the existence of a changepoint. If $Z_j = 1$, the model treats the ordinal predictor $x_{ij}$ as a continuous variable by assuming no threshold exists in the effect and instead there is an equal distance between two consecutive values. If $Z_j = 0$, a sudden jump occurs in the effect of the predictor at some value. To account for this change, the model treats the ordinal predictor as a binary variable, which is dichotomized into high and low values based on a threshold $\tau_j$. Through Bayesian modelling, parameters $Z_j$ and $\tau_j$ ($j = 1, \ldots, J$) are estimated from the data, and the posterior distributions of these parameters provide information on whether the ordinal predictor should be considered as a continuous or if a changepoint exists, where to dichotomize the predictor as the appropriate functional form. It is important to note that each predictor is divided by twice its standard deviation when $Z_j = 1$.[14] This step is critical for the model to successfully distinguish $Z_j = 1$ from $Z_j = 0$ by normalizing the continuous and binary variables on the same scale.

For prior specification, $Z_j$ is assumed to have a Bernoulli distribution with probability $p_z$, and parameter $\tau_j$ is assumed to have a discrete categorical distribution taking each value $k$ ($k \in \{1, ..., K\}$) with probability $\pi_k$. For other regression parameters $\alpha$ and $\beta_j$, we use normal priors that are common in standard linear models. Specifically, parameters $\alpha$ and $\beta_j$ have a normal distribution with zero mean and a large variance indicating vague information in the prior, and the residual variance, $\sigma^2$ has an inverse gamma distribution. To estimate the function of $x_j$, $p_z$ has a flat prior $p_z \sim Beta\ (0.5, 0.5)$, and $(\pi_1, \pi_2, \pi_3) \sim Dirichlet\ (1,1,1)$ in estimation of $\tau_j$.

If the outcome $y_i$ is binary, we assume that $y_i \sim Bernoulli(\mu_i)$ and model $y_i$ as a function of $f(x_{ij})$ through a logit or probit link, $g(*)$:

$$g(\mu_i) = \alpha + \beta_1 f(x_{i1}) + \cdots + \beta_j f(x_{ij}) + \cdots + \beta_J f(x_{iJ}), \quad (2)$$

If the outcome is time-to-event, we use a Cox proportional hazards regression. We formulate the Cox model using counting processes.[15] Let $N_i(t)$ be the number of events that occurred up to time $t$ and $dN_i(t)$ be the increment of $N_i$ over a small time interval [t, t+dt]. If subject $i$ experiences the event during this interval, $dN_i(t)$ will take the value 1; otherwise, $dN_i(t)$ is 0. The counting process increments $dN_i(t)$ are assumed to be independent Poisson random variables with means $d\lambda_i(t)$. For survival analysis, we model $d\lambda_i(t)$ as a function of ordinal predictors, $x_{ij}$ ($j = 1, ..., J$). Thus, the model is described by

$$dN_i(t) \sim Poisson(d\lambda_i(t)), \quad (3)$$

$$d\lambda_i(t) = Y_i(t) \exp\left(\beta_1 f(x_{i1}) + \cdots + \beta_j f(x_{ij}) + \cdots + \beta_J f(x_{iJ})\right) d\Lambda_0(t)$$

Here, $Y_i(t) = 1$ if subject $i$ is under observation at time $t$ and 0 otherwise. $d\Lambda_0(t)$ is the increment in the cumulative baseline hazard function during the time interval [t, t+dt].

The gamma process (GP) is used as a prior for $\Lambda_0(t)$,[16] that is, $\Lambda_0(t) \sim GP(c_0\widehat{\Lambda}_0, c_0)$, where $\widehat{\Lambda}_0$ is often assumed to be a known parametric function, and $c_0$ represents the degree of confidence in this prior.

We want to emphasize that our model is very easy to implement. Parameters were estimated in Just Another Gibbs Sampler, known as JAGS,[17-18] a statistical software package that uses Markov Chain Monte Carlo to generate samples from the relevant posterior distributions. The convergence of the Markov chain was assessed visually and with the Gelman-Rubin Diagnostic criterion.[19] JAGS code can be found in the Appendix (availiable upon request).

2.2 Penalized regression model

Our model (1)-(3) can be easily modified to situations where many predictors are present in the regression model. The Least Absolute Shrinkage and Selection Operator method, commonly known as Lasso, is often considered one standard method for variable selection.[20] By introducing a penalty term to shrink certain coefficient estimates toward zero, the method encourages a more simple, sparse solution. Lasso is commonly implemented to reduce a large set of potential biomarkers by estimating coefficients of non-important biomarkers as zero while allowing for large coefficients for relevant biomarkers.[21] The additional constraint contains a tuning parameter than controls the amount of shrinkage and therefore selection under the frequentist paradigm. In the Bayesian framework, the Laplace or equivalently double exponential prior corresponds to the lasso penalty.

Lasso prior:

$$\beta_j \sim DE(0, \lambda \times v)$$

$$v \sim gamma(0.01, 0.01)$$

Where DE is the double exponential distribution, $v = 1/\sigma^2$ is the precision of the regression model, and $\lambda$ is the shrinkage parameter which controls the prior variance and therefore different levels of shrinkage. Large values of $\lambda$ correspond to large variance and introduce low information in the Bayesian model by essentially imposing no shrinkage on the model parameters. On the other hand, small values of $\lambda$ correspond to strong prior that the vector $\beta$ is close to zero and result in high levels of shrinkage. We fix the value of $\lambda$ in the following simulation settings.

Some researchers have noted that the Laplace prior may shrink large coefficients more than is necessary. An alternative for conducting covariate selection is the horseshoe estimator based on a scale mixture normal model.[22] To avoid too many exact zero-estimates, this prior incorporates both a local and global shrinkage parameter for further flexibility of shrinking some coefficients to zero while leaving others to remain large.

Horseshoe prior:

$$\beta_j \sim N(0, \lambda \times v_s \times v_j)$$

$$v_j \sim Half - Cauchy(0, 1)$$

$$v_s \sim Half - Cauchy(0, 1)$$

The $\lambda$ has the same meaning as in the Lasso prior. JAGS code with the lasso and horseshoe prior can be found in the Appendix (available upon request).

## 3 SIMULATION STUDIES

For each type of outcome (continuous, binary, and survival), a series of simulation studies are conducted to evaluate the proposed method. Posterior inference is based on the parameters $\beta$ and $\tau$. We present the true coefficient $\beta$, simulation-based average posterior mean $\hat{\beta}$, standard deviation (SD), mean square error (MSE), and coverage probability (CP) of the 95% credible intervals containing the true $\beta$ in all simulation studies. Posterior probabilities of each $\tau$ cutoff value were collected from JAGS. A high posterior probability $P(Z_j = 1)$ supports $x_j$ having no changepoint and analyzing it as a numeric variable. If $P(Z_j = 1)$ is low, dichotomization of $x_j$ is preferred via a threshold at the estimated changepoint. Posterior probabilities $P(\tau_j = 1 | Z_j = 0)$, $P(\tau_j = 2 | Z_j = 0)$, and $P(\tau_j = 3 | Z_j = 0)$ are used to determine the best cutoff. The probabilities presented represent the proportion of simulations the model determined the variable was linear or binary at each value. All simulations were completed in R using the R2jags interface.[23-24]

### 3.1 Simulation Studies for Data with a Continuous Outcome

Patient data are generated for 35 subjects for the linear regression model. Within each simulation, 12 predictor values are generated from a normal distribution for each patient. For the results shown below that include implementation of the regularization methods, we generated 13 covariates with correlation 0, though we also conducted simulations without selection and high correlation with good performance. For the Lasso and Horseshoe prior, we set the tuning parameter $\lambda$ at 0.01. By categorizing these generated values at the 30$^{\text{th}}$, 60$^{\text{th}}$, and 85$^{\text{th}}$ percentiles, the data matrix $x$ is comprised of

4 ordinal values 0, 1, 2, and 3. A matrix of indicators is produced to denote which predictors are dichotomous at some threshold and which are continuous and should be scaled by two standard deviations. With true coefficient values fixed at 0, 1, 2, and 3, the outcome $y_i$ is generated from a normal distribution based on the data $x$ and coefficient $\beta$ with standard deviation 0.1. For each model, 200 simulations are conducted with JAGS with 10,000 iterations of 2 chains and a burn-in of 2,000 iterations using the priors described previously.

Linear regression results in Tables I and II show that with a small sample size and moderate correlation, the coefficients are accurately estimated. For the non-zero regression coefficients, the model correctly identifies the correct value of $\tau$ representing the true linearity or changepoint of the variable. Alternatively, the first row demonstrates that when the predictor has no true effect, the model does not conclusively decide if a changepoint exists in the null effect. Performance of the method is also good when using the Lasso and horseshoe penalties for covariate selection as shown in Tables I and II respectively. By including the posterior distribution of the inclusion indicator, we see that sufficiently important predictors are selected in the models. Overall, the coverage probabilities are relatively close to a desirable level of 0.95, and the model consistently select the covariates with a non-zero effect size.

*Table I. Parameter estimates and performance statistics of JAGS simulation for linear regression data using Lasso penalization method.*

| $\beta$ | $\hat{\beta}$ | SD | MSE | CP | Selection Proportion | Cutoff $\tau$ | Posterior Probabilities | | | |
|---|---|---|---|---|---|---|---|---|---|---|
| | | | | | | | $Z=1$ | $\tau_1$ | $\tau_2$ | $\tau_3$ |
| 0 | 0.007 | 0.046 | 0.002 | 0.990 | 0.151 | $\tau_1$ | 0.477 | 0.163 | 0.174 | 0.186 |
| 0.25 | 0.241 | 0.087 | 0.008 | 0.900 | 0.970 | None | 0.855 | 0.045 | 0.072 | 0.028 |
| 0.5 | 0.487 | 0.070 | 0.005 | 0.890 | 0.992 | None | 0.973 | 0.013 | 0.009 | 0.005 |
| 0 | -0.013 | 0.050 | 0.003 | 0.980 | 0.166 | None | 0.486 | 0.172 | 0.158 | 0.185 |
| 0.25 | 0.243 | 0.086 | 0.007 | 0.940 | 0.929 | $\tau_1$ | 0.138 | 0.833 | 0.016 | 0.013 |
| 0.5 | 0.497 | 0.072 | 0.005 | 0.910 | 0.992 | $\tau_1$ | 0.012 | 0.985 | 0.001 | 0.001 |
| 0 | 0.001 | 0.036 | 0.001 | 1.000 | 0.168 | $\tau_1$ | 0.493 | 0.166 | 0.159 | 0.182 |
| 0.25 | 0.237 | 0.056 | 0.003 | 0.930 | 0.974 | $\tau_2$ | 0.233 | 0.006 | 0.753 | 0.009 |
| 0.5 | 0.497 | 0.075 | 0.006 | 0.890 | 0.989 | $\tau_2$ | 0.048 | 0.002 | 0.947 | 0.002 |
| 0 | -0.002 | 0.070 | 0.005 | 1.000 | 0.178 | $\tau_2$ | 0.472 | 0.166 | 0.165 | 0.197 |
| 0.25 | 0.191 | 0.132 | 0.021 | 0.960 | 0.814 | $\tau_3$ | 0.182 | 0.034 | 0.043 | 0.741 |
| 0.5 | 0.465 | 0.113 | 0.014 | 0.890 | 0.964 | $\tau_3$ | 0.040 | 0.007 | 0.007 | 0.947 |
| 0 | 0.002 | 0.041 | 0.002 | 1.000 | 0.162 | $\tau_3$ | 0.485 | 0.167 | 0.168 | 0.181 |

*Table II. Parameter estimates and performance statistics of JAGS simulation for linear regression data using the horseshoe prior penalization method.*

| $\beta$ | $\hat{\beta}$ | SD | MSE | CP | Selection Proportion | Cutoff $\tau$ | Posterior Probabilities | | | |
|---|---|---|---|---|---|---|---|---|---|---|
| | | | | | | | $Z=1$ | $\tau_1$ | $\tau_2$ | $\tau_3$ |
| 0 | -0.001 | 0.025 | 0.001 | 1.000 | 0.192 | $\tau_1$ | 0.489 | 0.174 | 0.162 | 0.175 |
| 0.25 | 0.247 | 0.062 | 0.004 | 0.960 | 0.960 | None | 0.841 | 0.031 | 0.076 | 0.051 |
| 0.5 | 0.494 | 0.051 | 0.003 | 0.940 | 0.999 | None | 0.980 | 0.001 | 0.016 | 0.003 |
| 0 | 0.000 | 0.027 | 0.001 | 1.000 | 0.179 | None | 0.488 | 0.169 | 0.166 | 0.177 |
| 0.25 | 0.218 | 0.068 | 0.006 | 0.980 | 0.910 | $\tau_1$ | 0.194 | 0.758 | 0.026 | 0.021 |
| 0.5 | 0.488 | 0.052 | 0.003 | 0.990 | 0.997 | $\tau_1$ | 0.034 | 0.963 | 0.002 | 0.001 |
| 0 | 0.001 | 0.023 | 0.001 | 1.000 | 0.182 | $\tau_1$ | 0.490 | 0.166 | 0.167 | 0.177 |
| 0.25 | 0.226 | 0.063 | 0.004 | 0.970 | 0.930 | $\tau_2$ | 0.255 | 0.018 | 0.701 | 0.026 |
| 0.5 | 0.481 | 0.052 | 0.003 | 0.990 | 0.997 | $\tau_2$ | 0.045 | 0.001 | 0.952 | 0.001 |
| 0 | -0.001 | 0.023 | 0.001 | 1.000 | 0.177 | $\tau_2$ | 0.489 | 0.166 | 0.167 | 0.177 |
| 0.25 | 0.169 | 0.099 | 0.016 | 1.000 | 0.730 | $\tau_3$ | 0.253 | 0.056 | 0.066 | 0.624 |
| 0.5 | 0.456 | 0.104 | 0.013 | 0.990 | 0.960 | $\tau_3$ | 0.052 | 0.009 | 0.012 | 0.927 |
| 0 | 0.004 | 0.032 | 0.001 | 1.000 | 0.181 | $\tau_3$ | 0.491 | 0.167 | 0.167 | 0.176 |

### 3.2 Simulation Studies for Data with a Binary Outcome

Patient data are generated for 350 subjects for the logistic regression setting to account for the lesser degree of information provided by binary outcomes. The predictor values are simulated in the same way as the linear regression model with correlation 0.25 among subject predictors and corresponding outcomes sampled from a binomial distribution with inverse logit values for probabilities.

*Table III. Parameter estimates and performance statistics of JAGS simulation for logistic regression data.*

| $\beta$ | $\hat{\beta}$ | SD | MSE | CP | Cutoff $\tau$ | Posterior Probabilities | | | |
|---|---|---|---|---|---|---|---|---|---|
| | | | | | | $Z=1$ | $\tau_1$ | $\tau_2$ | $\tau_3$ |
| 0 | 0.003 | 0.480 | 0.231 | 0.920 | None | 0.441 | 0.165 | 0.167 | 0.227 |
| 3 | 3.203 | 0.454 | 0.247 | 0.945 | None | 0.984 | 0.000 | 0.016 | 0.000 |
| 2 | 2.108 | 0.494 | 0.256 | 0.915 | $\tau_1$ | 0.087 | 0.910 | 0.002 | 0.002 |
| 3 | 3.220 | 0.399 | 0.208 | 0.940 | $\tau_2$ | 0.030 | 0.000 | 0.970 | 0.000 |
| 2 | 2.121 | 0.715 | 0.526 | 0.890 | $\tau_3$ | 0.105 | 0.006 | 0.016 | 0.873 |

For logistic regression with no penalization, the estimated values of $\hat{\beta}$ in Table III are close to the true values of $\beta$, but more bias is introduced when the outcome is binary. We see the model is able to identify the true threshold cutoff level $\tau$ for each binary predictor by allotting the most posterior probability to the appropriate changepoint. When a consensus is not reached for a majority posterior probability, it is under the setting of the null effect of the predictor. Additionally, the simulations have slight under-coverage of the true $\beta$ being in the 95% credible interval.

## 3.3 Simulation Studies for Data with a Survival Outcome

Patient data are generated for 150 subjects as described above. The survival times for the Cox Proportional Hazards model are generated from an exponential distribution with a hazard of 1, and the censoring proportion is simulated from an exponential model with parameter 0.1. For survival data priors, the mean of the gamma distribution represents the anticipated value or mean of the hazard function, with a small value of $c$ indicating a weak prior belief. For this method, we use $c = 0.001$, $r = 0.01$, and $dL_{0j}^* = r \cdot (t(j+1) - t(j))$.

Table IV. Parameter estimates and performance statistics of JAGS simulation for survival data regression model.

| $\beta$ | $\hat{\beta}$ | SD | MSE | CP | Cutoff $\tau$ | Posterior Probabilities | | | |
|---|---|---|---|---|---|---|---|---|---|
| | | | | | | $Z=1$ | $\tau_1$ | $\tau_2$ | $\tau_3$ |
| 0 | 0.011 | 0.26 | 0.068 | 0.950 | None | 0.421 | 0.172 | 0.171 | 0.236 |
| 3 | 3.142 | 0.294 | 0.107 | 0.905 | None | 0.999 | 0.000 | 0.000 | 0.001 |
| 2 | 2.099 | 0.294 | 0.096 | 0.925 | $\tau_1$ | 0.000 | 1.000 | 0.000 | 0.000 |
| 3 | 3.131 | 0.285 | 0.098 | 0.935 | $\tau_2$ | 0.000 | 0.000 | 1.000 | 0.000 |
| 2 | 2.043 | 0.307 | 0.096 | 0.935 | $\tau_3$ | 0.007 | 0.000 | 0.000 | 0.993 |

In this survival simulation, the average censoring proportion is 0.199. We see in Table IV that the model can identify the true threshold level, $\tau$, in each case where the true value of $Z$ is 0, meaning a changepoint exists. In these cases, the true value of $\tau$ has the greatest posterior probability. While the coverage probabilities of the true value being in the 95% coverage interval are not at the nominal 95% level, the credible interval contains the true value of $\beta$ in at least 90% of the simulations.

## 4. APPLICATIONS

For binary data, we apply our method to a dataset of 121 observations from fine needle aspirations and 21 ordinal cytologic markers that have been considered elsewhere.[25] Since some biomarkers are highly correlated, and the selection methods we consider here are prone to worse performance with correlated data, we do not consider variable selection for these data with only moderate sample size. We evaluate the significance of the markers based on the value 0 not being within the 95% credible intervals. The model allows for the number of possible cutoffs to vary for each variable since different markers have different numbers of possible categories. A frequency table was used to decide the potential number of changepoint values based on the sample size in each to ensure any possible threshold value had sufficient data available for estimation. For example, a covariate with a low frequency in the third ordered level would have only two potential cutoff values, while those with sufficient data for all four levels have three potential cutoff values. The prior parameterizations were the same as for the simulation studies above.

Table V. Significant multivariate logistic regression results of proposed method on fine needle aspiration data using 21 predictors.

| Marker | $\hat{\beta}$ | 95% CI $\hat{\beta}$ | Posterior Probabilities | | | |
|---|---|---|---|---|---|---|
| | | | $Z = 1$ | $\tau_1$ | $\tau_2$ | $\tau_3$ |
| Syncytial tissue fragments | 4.245 | **(1.062, 7.509)** | 0.141 | 0.003 | **0.852** | 0.004 |
| Microfollicles within tissue fragments | -2.744 | **(-5.204, -0.356)** | 0.370 | 0.469 | 0.161 | NA |
| Isolated microfollicles | 2.393 | **(0.046, 4.750)** | **0.763** | 0.093 | 0.144 | NA |
| Number overlapping | -4.278 | **(-8.493, -0.428)** | 0.360 | 0.021 | 0.007 | 0.612 |
| Number with enlargement | -4.961 | **(-9.989, -0.176)** | 0.184 | 0.021 | **0.795** | NA |
| Coarse chromatin | 6.019 | **(0.695, 13.127)** | 0.471 | 0.016 | 0.513 | NA |
| Colloid | -1.709 | **(-3.358, -0.100)** | 0.400 | 0.195 | 0.405 | NA |

Our method found several variables which produced a credible interval that does not contain 0, shown in Table V. When examining which changepoint value based on $\tau$ has the greatest posterior probability for these significant characteristics, we see that the cytomorphologic characteristic of isolated microfollicle cells should be treated as a linear variable based on the high posterior probability of $Z = 1$. Based on this sample size, two variables seem to have a changepoint effect and reach consensus for an appropriate threshold to dichotomize the predictor. The number of cells with nuclear enlargement and architecture trait of syncytial tissue fragments both have the greatest posterior probability for $\tau_3$, suggesting a changepoint of two may be used for those variables.

Other variables have posterior probabilities that do not decidedly suggest whether a changepoint exists.

To evaluate the method on survival data, a dataset containing low grade glioma patients was obtained from The Cancer Genome Atlas (TCGA).[26] The final dataset contained 396 patients with time-to-overall survival information and a mutation in the main genetic lesion for these patients known as LGG-mIDH1. After a univariate preliminary screening, we considered 45 gene expression values as predictors. The gene measurements were categorized into ordered quantiles and entered into the model. To perform variable selection, we incorporated the LASSO penalty with a tuning parameter value of $\lambda = 0.001$.

Table VI. Significant multivariate survival data regression results of proposed method on TCGA data with LASSO variable selection from 45 predictors.

| Markers | $\hat{\beta}$ | 95% CI $\hat{\beta}$ | Posterior Probabilities | | | |
|---|---|---|---|---|---|---|
| | | | $Z = 1$ | $\tau_1$ | $\tau_2$ | $\tau_3$ |
| ACRV1 | 1.320 | **(0.688, 1.917)** | 0.001 | 0.001 | 0.002 | **0.996** |
| CDKN2B | -4.625 | **(-7.011, -1.727)** | 0.001 | **0.997** | 0.001 | 0.001 |

Among the 45 predictors considered, the method identified a significant relationship with both ACRV1 and CDKN2B and overall survival time with results shown in Table VI. We further assessed the posterior probabilities for $\tau$ to determine that changepoints may exist in the effects of these markers. The model shows ARCV1 may be have a threshold based on the highest quartile, while CDKN2B suggests a cutoff of being in or above the lowest quartile as a dichotomization point.

## 5. DISCUSSION

This paper proposes a novel Bayesian approach to perform regression modeling and determine if an ordinal predictor should be dichotomized, and if so, at what changepoint value. Regression coefficients for linear, logistic, and survival models are estimated with this model, and posterior probabilities are assessed to determine an appropriate cut point value. Simulations suggest that with a moderate sample size, the various regression models perform well. The model is accurately able to determine the true form and value of the regression coefficient. Simulations show that the coverage in some scenarios was less than the nominal rate, which may be due to the fact that the regression coefficient and corresponding credible intervals are averaged over all levels of $\tau$, when in truth the binary variable distribution comes from only one level of $\tau$. Applied to clinical settings, the real data sets demonstrate the ability of the model to identify significant markers related to both logistic and survival outcomes. In addition, the model can determine which biomarkers are most likely to have a changepoint effect and which level to use as the threshold.

A strength of the method is its ability to estimate the coefficient value of a variable while at the same time determining what form it should take based on a changepoint. If the model returns relatively equal posterior probabilities of threshold values of $\tau$, interpretation of the best threshold may be limited as we do not recommend an objective criterion to choose such a cut point. It is possible that all ordinal levels appear equally likely to be a cutoff point, which may be evidence to treat each level as a category in the model. Alternatively, simulations suggest this ambiguity may stem from a predictor with no true effect. In that case, the variable may not be sufficiently important to include in the model or decision-making. By using a Bayesian framework, this method naturally considers the uncertainty of estimating the changepoint and provides a distribution for the threshold. Further, the model can easily incorporate existing information about a biomarker or prior knowledge about a clinically valid threshold point. Particularly for biomarkers that have previously contradictory thresholds for decisions in the field, wagering this information through a prior distribution and the data may yield more conclusive decision rules.

While our method is intended to be useful in a setting where a clinician must make some decision for a patient, other situations may warrant deciding if a variable should be treated as linear or considered as multiple categories. In future work, the model could be extended to combine categories to estimate more than one cutoff value and could also be extended beyond linear, binary, and survival outcomes to relax distributional assumptions. One limitation of the method is that it does not directly account for the correlation between covariates during the estimation procedure. When patient characteristics are strongly correlated in data applications, the determination of which variable to dichotomize and which to treat as linear in the model may be less clear. This potential difficulty may be likened to the challenge of variable selection for highly correlated variables. We note that our method is amenable to other variable selection methods that extend beyond LASSO and the horseshoe prior considered here.

Overall, the proposed method presents a useful option to determine if an ordinal variable should be included in a model by using standardization to estimate the linear or changepoint-based form of the predictor. By simultaneously estimating the regression coefficient and credible interval, this framework is a more rigorous method to fit a regression model with ordinal predictors. The method is practical with its use of JAGS to obtain posterior distributions to conduct inference for various types of outcomes and can be implemented in both medical and more general settings.

**Declaration of conflicting interests**

The Authors declare that there is no conflict of interest.